\documentclass[aps,prb,reprint,amssymb,superscriptaddress,showpacs]{revtex4-1}

\usepackage{rotating}
\usepackage{amsmath}
\usepackage{amssymb}
\usepackage{color}
\usepackage{graphicx}
\usepackage[active]{srcltx}
\usepackage[sort&compress]{natbib}
\usepackage[dvips]{epsfig}
\usepackage{float}
\usepackage{comment}

\renewcommand{\[}{\left[}
\renewcommand{\]}{\right]}
\renewcommand{\(}{\left(}
\renewcommand{\)}{\right)}

\newcommand{\yambo} {{\normalfont\ttfamily YAMBO }}

\newcommand{\pwscf} {{\normalfont\ttfamily PWSCF }}

\newcommand{\be}{\begin{equation}}
\newcommand{\ee}{\end{equation}}
\newcommand{\bea}{\begin{eqnarray}}
\newcommand{\eea}{\end{eqnarray}}
\newcommand{\bml}{\begin{multline}}
\newcommand{\eml}{\end{multline}}

\def\ai		{{\em ab--initio}}

\def\rr		{{\bf r}}
\def\xx		{{\bf x}}

\def\GG		{{\bf G}}

\def\RR		{{\bf R}}

\def\PP		{{\bf P}}

\def\qq		{{\bf q}}
\def\zero {{\bf 0}}
\def\kk		{{\bf k}}

\def\ra         {\rangle}

\renewcommand{\[}{\left[}
\renewcommand{\]}{\right]}
\renewcommand{\(}{\left(}
\renewcommand{\)}{\right)}

\def\nk         {n{\bf k}}

\def\dk         {\frac{d\,{\bf k}}{\(2 \pi\)^3}}
\def\dk1        {\frac{d\,{\bf k}_1}{\(2 \pi\)^3}}

\def\dk         {\frac{d\,{\bf k}}{\(2 \pi\)^3}}

\def\nk         {n{\bf k}}

\def\bverb      {\renewcommand{\baselinestretch}{1}\begin{verbatim}}
\def\everb  	{\end{verbatim}\renewcommand{\baselinestretch}{1.3}}

\def\g{\gamma}
\def\G{\Gamma}
\def\d{\delta}
\def\D{\Delta}
\def\e{\epsilon}
\def\ve{\varepsilon}
\def\vet{\tilde{\varepsilon}}
\def\h{\eta}

\def\l{\lambda}

\def\PP{\boldsymbol {\cal P}}
\def\efield{\boldsymbol{\cal E}}
\def\r{\rho}

\def\Rt{\tilde{R}}
\def\Lt{\tilde{L}}
\def\Et{\tilde{E}}
\def\et{\tilde{\epsilon}}
\def\ft{\tilde{f}}
\def\Ht{\tilde{H}}
\def\At{\tilde{A}}

\def\S{\Sigma}
\def\St{\tilde{\Sigma}}
\def\t{\tau}

\def\x{\chi}
\def\xt{\tilde{\chi}}

\def\w{\omega}
\def\W{\Omega}

\def\ra{\rightarrow}

\def\Re{{\rm Re}}
\def\Im{{\rm Im}}

\def\1op{\hat{\mathbbm{1}}}
\def\1{\mathbbm{1}}

\newcommand{\cnrmlb} {Istituto di Struttura della Materia of the National Research Council --
                      Via Salaria Km 29.3, 00016 Monterotondo Stazione, Italy}
\newcommand{\etsf} {European Theoretical Spectroscopy Facilities (ETSF)}
\newcommand{\polimi} {Dipartimento di Fisica, Politecnico di Milano and
                      Institute of Photonics and Nanotechnologies, IFN-CNR -- Piazza L. da Vinci 32, 20133 Milano, Italy}

\begin{document}
 
\title{Non equilibrium optical properties in semiconductors from first--principles:
      a combined theoretical and experimental study of bulk silicon}

\author{Davide Sangalli}
\affiliation{\cnrmlb} 
\affiliation{\etsf} 

\author{Stefano Dal Conte}
\affiliation{\polimi} 

\author{Cristian Manzoni}
\affiliation{\polimi} 

\author{Giulio Cerullo}
\affiliation{\polimi} 

\author{Andrea Marini}
\affiliation{\cnrmlb} 
\affiliation{\etsf} 

\date{\today}
\begin{abstract}
The calculation of the equilibrium optical properties of bulk silicon by using the Bethe--Salpeter
equation solved in the Kohn--Sham basis represents a cornerstone in the development of
an {\em ab--initio} approach to the optical and electronic properties of materials.
Nevertheless calculations of the {\em transient} optical spectrum
using the same efficient and successful scheme are scarce.
We report, here, a joint theoretical and experimental study of the
transient reflectivity spectrum of bulk silicon.
Femtosecond transient reflectivity is compared to a
parameter--free calculation based on the non--equilibrium
Bethe--Salpeter equation. By providing an accurate description of the experimental results
we disclose the different phenomena that determine the transient optical response
of a semiconductor.
We give a parameter--free interpretation of concepts
like bleaching, photo--induced absorption and stimulated emission, beyond
the Fermi golden rule. We also introduce the concept of optical gap renormalization,
as a generalization of the known mechanism of band gap renormalization.
The present scheme successfully describes the
case of bulk silicon, showing its universality and accuracy.
\end{abstract}
\maketitle

\section{Introduction}
Understanding the electronic and optical properties of molecules and solids
requires knowledge of their structure and their dynamics. Structural
information describes the time-independent relative arrangement
of the atoms in the system and links their physical properties
to their spatial location at equilibrium. Dynamics describes how the system
responds to external stimuli and evolves from one configuration to
the other when driven in the non--equilibrium (NEQ) regime.
Many dynamical processes, involving for example the motion of atoms
or charge carriers occur on very short time-scales, ranging from
$10^{-14}$ to $10^{-12}$ s. 
Informations on such ultrafast events can be gained by ultrafast optical spectroscopy~\cite{Mukamel1999},
a technique which investigates time--dependent variations of the optical properties of a
system under the effect of an ultrashort light pulse excitation. Since the availability of
femtosecond laser pulses in the early 80s, there has been an enormous progress in
this field. In particular some key technological advances
in the 90s, such as mode-locked Ti:sapphire lasers, chirped pulse
amplification~\cite{Backus1998} and optical parametric amplifiers~\cite{Brida2010}
have greatly improved the stability and reliability of ultrafast laser systems, making
femtosecond spectroscopy available to a broad community of non-specialist users. 

Several experimental methods have been developed for solids, starting from the simple
pump-probe, which allows to measure photoinduced population dynamics by monitoring the
time-dependent transmission or reflectivity of the sample with time resolution down to
a few light cycles~\cite{Knox1986,Leitenstorfer1996}. Technical advances have enabled
more sophisticated techniques, from photon echo~\cite{Becker1988} and two-dimensional~\cite{Stone2009} 
spectroscopies, which measure the polarization dephasing times and the correlations
between photo--excitations, to time- and angle--resolved photo--electron spectroscopy~\cite{Bovensiepen2012},
tracking time- and momentum-dependent electron energy distributions~\cite{Schmitt2008}, to time-resolved
$THz$ spectroscopy~\cite{Ulbricht2011}, which monitors the non-equilibrium evolution of charge
carriers and low-energy excitations~\cite{Huber2001}.
Extensive experiments have been performed on bulk and nanostructured semiconductors~\cite{Shah1999},
superconductors~\cite{DalConte2015b} and other strongly correlated materials~\cite{Basov2011}.

The situation from the theoretical and computational point of view is rather different.
Indeed, the modeling of equilibrium and out--of--equilibrium
materials has followed several distinct and often fragmented and uncorrelated paths.
The most up--to--date scheme to calculate and predict the equilibrium ground-- and excited--state
properties of a wide range of materials is based on the merging of
Density--Functional--Theory~\cite{R.M.Dreizler1990} (DFT), with
Many--Body Perturbation Theory~\cite{Onida2002} (MBPT). 
Such approach is often referred to as \textit{ab-initio} Many-Body
Perturbation Theory~\cite{Onida2002} (\textit{ai}--MBPT).
Within \textit{ai}--MBPT, DFT provides a suitable single--particle basis
for the MBPT scheme.
This methods has been applied successfully, for example, to correct the well--known
band--gap underestimation problem of DFT~\cite{GW_review,VanSchilfgaarde2006},
and to calculate the optical properties of an impressive range of materials~\cite{Onida2002}.

The \textit{ai}--MBPT has been constantly developed in the last decades leading to a solid framework.
Already in the seminal work of Hanke and Sham~\cite{Hanke1980}, it was clear that only
by properly including the combined effect of the screened electron--electron and electron--hole 
interactions, it is possible to correctly interpret the equilibrium absorption spectrum
of extended systems in terms of excitonic states.
The electron--electron interaction is included in the computation of the
quasi--particles (QP) band structure by means of the $GW$ approximation,
while the electron--hole interaction is included
in the computation of the neutral excitations by means of the Bethe--Salpeter equation (BSE).
Therefore, as far as equilibrium properties are concerned, a number of well established standards
and codes are available to describe the experimental absorption spectra.

The situation in the out--of--equilibrium case is rather different.
The NEQ Green function
theory~\cite{Rossi2002,Koch2006,PhysRevB.38.3342},
\emph{i.e.} the NEQ version of MBPT, has only be used to simulate the
non-equilibrium response of simple ideal
materials~\cite{HartmutHaug2008,Bonitz1998,Stefanucci2013,Kadanoff1962,Perfetto2015b} while,
for more realistic materials, semi--empirical methods based on the semiconductor Bloch
equations~\cite{Fehrenbach1982,Collet1994,Glezer1995} and semi--classical Boltzmann
equation~\cite{Brida2013} have been adopted. 
As a consequence, while the absorption spectrum of bulk silicon, for example, is well
understood~\cite{Hanke1980,Lautenschlager1987,Albrecht1998,Marini2008,Onida2002},
its NEQ behavior is not.
It is then clear that there is a gap between the experimental and computational
approaches. The problem is, how to export to the transient case the same level of accuracy
and predictivity gained in the equilibrium limit by extending both the
BSE and the $GW$ approximation out--of--equilibrium. 

Only very recently a series of works have devised a formal extensions of
\textit{ai}--MBPT to the different aspects of the NEQ regime: 
the interaction with a laser pulse and the description of excitons in a real--time
fashion~\cite{Attaccalite2011}, the introduction, \ai, of the
electron--phonon and electron--electron scatterings in the carrier dynamics~\cite{Marini2013,Sangalli2015a},
the introduction of the NEQ Bethe--Salpeter equation~\cite{Perfetto2015a},
the derivation of a coherent and complete extension of Hedin's equations to the
out--of--equilibrium dynamics of electrons, phonons and photons~\cite{Melo2016} using the Kohn--Sham basis.
These methods have been applied successfully to study carrier
dynamics~\cite{Sangalli2015b,Bernardi2015,Bernardi2014,Sato2014}
and transient absorption~\cite{Pogna2016} in paradigmatic materials. 

However, the \ai\, way to describe out--of--equilibrium properties is
just moving the first steps and, as a consequence, it represents a rapidly
changing scenario. 
A winning scheme to get a real benchmark for theories and approximations
is to perform combined theoretical and experimental studies where the
simulations are on--the--fly compared with the measured quantities.
Indeed the ultrafast dynamics of semiconductors, largely explored by
spectroscopic measurements in the last decades, are among the first cases studied
theoretically and provided an excellent test--bed for the earlier development
of the theory.

In this work bulk silicon is used as a natural test--bed to the further development
and validation of the \textit{ai}--MBPT in the NEQ regime.
We will perform Transient Reflection (TR) measurements, in which the system
is excited by an ultrashort laser pulse, and its subsequent temporal evolution is measured
by detecting the reflection of a delayed broadband probe pulse.
We will compare the measured
out-of-equilibrium behavior of bulk silicon after photoexcitation with
state--of--the--art simulations. In order to simplify the analysis, 
we will pump the system at photon energies close to the optical gap.
In addition, we will work in the low density regime~\cite{HartmutHaug2008},
where the effect of the pump pulse can be described in terms of electrons
jumping from the occupied to the unoccupied orbitals with (small)
variation of the electronic population.

By comparing the simulations with the experiment, we will show that
\textit{ai}--MBPT in the NEQ regime is able to give a clear and complete
description of the non--equilibrium optical properties of bulk silicon.
We will give a formal definition of
bleaching, photo--induced absorption and stimulated emission. This will
make possible to associate these concepts, which otherwise can be hardly
identified in the spectrum of extended systems, to specific features of
the experimental signal. Moreover we will introduce the concept of optical--gap
renormalization, as a generalization of the well known concept of band--gap renormalization.
The latter is a variation of the band structure of
the system induced by the renormalization of the electron--electron interaction.
The former, which is also a consequence of the band--gap renormalization,
is additionally influenced by the renormalization of the electron--hole interaction.

The transient reflectivity spectrum of silicon will be described in terms of the sum of
two effects: (i) a \emph{bleaching} of the first absorption peak plus
(ii) a shift in the position and a reduction in intensity of the same
peak due to the \emph{optical--gap renormalization}. 
The bleaching will give a reduction of the reflectivity and thus will describe
the negative signal in the low energy part of the spectrum. On the other hand
the optical gap renormalization will describe the main feature in the experimental 
signal around $3.3\ eV$.

The structure of the paper is as follows.
Sec.~\ref{sec:eq_props} presents the equilibrium simulations introducing key
concepts like the $GW$ approximation and the Bethe--Salpeter equation.
In Sec.~\ref{sec:trabs} we will introduce the NEQ optical properties of bulk silicon.
The experimental technique and results will be presented in sec.~\ref{subsec:TrAbs_exp}.
The theoretical method and the results of the simulations will be discussed in the subsequent sections
as follow.
In sec.~\ref{subsec:neq_dyn} the modeling of the pump pulse and the creation
of the NEQ population will be presented.
In sec.~\ref{subsec:TrAbs_theo_residuals} we will introduce a formal definition of
bleaching, photo--induced absorption and stimulated emission resulting from the NEQ population.
In sec.~\ref{subsec:TrAbs_theo_optical_gap} we will define the optical gap renormalization.
Finally in sec.~\ref{subsec:TrAbs_theo} we will give the complete description of the experimental
TR signal of silicon using the concepts introduced in the previous sections.

\section{Equilibrium properties}
\label{sec:eq_props}
Bulk silicon is a prototype semiconductor whose optical properties have been extensively studied
both from the experimental~\cite{Lautenschlager1987,Johansson1990} and the theoretical
side\cite{Hanke1980,Albrecht1998,Marini2008,Onida2002,Olevano2001}.
Here we start by describing its equilibrium electronic and optical properties showing that the
\textit{ai}--MBPT approach successfully describes the experimental results.

We start from a density--functional theory (DFT) calculation~\cite{Hohenberg1964}. 
Then we apply MBPT to obtain the
QP energies within the GW approximation~\cite{Hedin1965,Ferdi1998} and the absorption
spectra from the solution of the BSE~\cite{Onida2002}.
The QP band structure will show the effect of the screened
electron--electron interaction, while the absorption spectrum within
the BSE accounts for the effect of the electron--hole interaction.

The calculations of the equilibrium optical and electronic properties represents
a necessary basis for the NEQ simulations described in the next section.

\subsection{Quasi--particles and the electron--electron interaction}
\label{sec:eq_props:QP}
The Kohn--Sham (KS) ground--state~\cite{Kohn1965} is computed,
within the local density approximation~\cite{Perdew1981},
by using the \pwscf code~\cite{Giannozzi2009}.
The parameters for the calculations are a $4x4x4$ sampling $\kk$--grid
of the Brillouin zone, corresponding to a total of $64$ kpoints,
with an energy cut--off of $20\ Ry$ for the representation of the
wave--functions in reciprocal space.
The crystal structure is relaxed to its lowest energy and equilibrium configuration and norm conserving
pseudo--potentials are used to describe the $1s$, $2s$ and $2p$ electrons, which are kept
frozen in the core. Two atoms are present in the unit cell and thus there
are in total 4 occupied bands in the ground--state with 8 electrons
per unit cell, for which the $KS$ wave--functions, $\psi^{KS}$, and
eigen--values, $\et^{KS}$, are obtained.

In order to simplify the notation we represent in the following as $\tilde{O}$ a quantity $O$ evaluated at equilibrium.

MBPT corrections are then computed using the \yambo code~\cite{Marini2009}.
The QP energies $\et_{\nk}$ are defined, starting from the
KS energies, via the QP equation
\be
\et_{\nk}= \et^{KS}_{\nk}+\langle\nk|\St^{GW}(\et_{\nk})-\tilde{V}^{xc}|\nk\rangle
\text{,}
\label{eq:QP_GW_eq}
\ee
where $n\kk$ are indexes for the band\,($n$) and the k--point\,($\kk$).
$\langle\nk|\St^{GW}(\et_{\nk})|\nk\rangle$ is the expectation value of
the standard $GW$ self--energy~\cite{Hedin1965,Ferdi1998} 
${\St^{GW}(\xx,\xx';\w)}$ evaluated at $\w=\e_{n\kk}$,
where $\xx$ is a spatial coordinate and $\w$ is a frequency.
Similarly $\langle\nk| \tilde{V}^{xc}|\nk\rangle$ is the average of
${\tilde{V}^{xc}(\xx)\d(\xx-\xx')}$, the exchange--correlation potential
used to compute the DFT ground--state.

Eq.~\ref{eq:QP_GW_eq} is solved on a $16x16x16$ $\kk$--grid in the Brillouin zone,
with the $KS$ wave--functions and energies computed from a non--self consistent
DFT calculation starting from a $4x4x4$ $\kk$--grid.

The $GW$ self--energy
\begin{multline}
\St^{GW}(\xx,\xx';\w)=\int \frac{d^3\kk}{8\pi^3}\ e^{i\kk(\RR-\RR')}
                      \int \frac{d^3\qq}{8\pi^3} \\ \int \frac{d\w'}{2\pi}
               \tilde{G}_{\kk+\qq}(\rr,\rr';\w+\w') \tilde{W}_{\qq}(\rr,\rr';\w'),
\label{eq:GW_eq}
\end{multline}
is constructed starting from the $KS$ Green function $\tilde{G}$ and the screened
electron--electron interaction $\tilde{W}$. Here $\xx=\RR+\rr$ with $\rr$ inside
the unit cell and $\RR$ the coordinate which identifies the position of each
unit cell in the whole space.

The $KS$ Green function has the form
\begin{multline}
\tilde{G}_{\kk}(\rr,\rr';\w)=2\sum_n \(\psi^{KS}_{n\kk}(\rr)\)^*\psi^{KS}_{n\kk}(\rr')\times \\
 \times\Big(\frac{\ft_{n\kk}}{\w-\et_{n\kk}^{KS}+i\eta/2}+ \frac{1-\ft_{n\kk}}{\w-\et_{n\kk}^{KS}-i\eta/2}\big),
\label{eq:G_KS}
\end{multline}
with $\ft_{n\kk}$ the equilibrium occupation factors. The prefactor $2$
is due to spin degeneracies since we are considering a non polarized
system. The sum over $n$ is performed considering 50 bands.

The screened electron--electron interaction is: 
\be
\tilde{W}_{\GG\GG'}(\qq,\w)=4\pi\, \sum_{\GG\GG'} \frac{\tilde{\ve}_{\GG\GG'}^{-1}(\qq,\w)}{|\qq+\GG|\ |\qq+\GG'|}
\label{eq:W_GG}
\text{,}
\ee
with its Fourier transform in reciprocal space
\be
\tilde{W}_{\qq}(\rr,\rr';\w')=\frac{\W}{N_\GG^2}\sum_{\GG\GG'} e^{i\GG\cdot\rr} \tilde{W}_{\GG\GG'}(\qq,\w') e^{-i\GG'\cdot\rr'}.
\label{eq:W_rr}
\ee
$\W=8\pi^3/V$ with $V$ the volume of the unit cell.
We use the microscopic random--phase approximation (RPA) for the screening
\be
\tilde{\ve}^{-1}_{\GG\GG'}(\qq,\w)=\delta_{\GG,\GG'}+\frac{4\pi\xt^{RPA}_{\GG\GG'}(\qq,\w)}{|\qq+\GG||\qq+\GG'|}
\text{,}
\label{eq:eps_micro_RPA}
\ee
whose $\GG=\GG'=\qq=0$ component is defined via the $\lim_{\qq\ra 0}$.
$\x^{RPA}$ is obtained solving the Dyson equation for the response function
\begin{multline}
\xt^{RPA}_{\GG\GG'}(\qq,\w)=\xt^{KS}_{\GG\GG'}(\qq,\w)+ \\
 \sum_{\GG''} \xt^{KS}_{\GG\GG''}(\qq,\w)v_{\GG''}(\qq)\xt^{RPA}_{\GG''\GG'}(\qq,\w)
\label{eq:chi_RPA_dyson}
\text{.}
\end{multline}
In Eqs.~\ref{eq:GW_eq}-\ref{eq:chi_RPA_dyson}
we used $59\ \GG$-vectors (which correspond to a kinetic energy
of $4.6\ Ry$) and thus $59\ \rr$-vectors in Eq.~\ref{eq:W_rr}.
Eq.~\ref{eq:chi_RPA_dyson} is solved in the static case $\w=0$ and
for the imaginary energy $\w=i\,27.2\ eV$, the frequency dependence of $W(\w)$ is then
obtained via a plasmon pole model~\cite{Godby1989}.

The $KS$ response function can be constructed from the $KS$ wave--functions and energies.
Here we define the oscillator strengths as
\be
\r_{nm\kk,\GG}(\qq)= \int d^3\rr\ \(u^{KS}_{n\kk}(\rr)\)^* e^{iG\cdot\rr}  u^{KS}_{m\kk+\qq}(\rr)
\text{,}
\ee
with $u^{KS}_{n\kk}(\rr)$ the periodic part of the $KS$ block wave--functions.
We now introduce the compact notation ${l=\{nm\kk\}}$ for the
generalized electron--hole indexes with
\bea
&&\tilde{f}_{l,\qq}        = \tilde{f}_{m\kk+\qq}-\tilde{f}_{n\kk}  \text{,}  \\
&&\et^{KS}_{l,\qq}  = \et^{KS}_{n\kk}-\et^{KS}_{m\kk+\qq} \text{,}
\label{eq:delta_E_f}
\eea
and
\bea
&&\Lt_{ll'}^{KS}(\qq,\w)=\frac{\d_{l,l'} }{\w-\et^{KS}_{l,\qq}+i\h} \label{eq:L_KS}  \text{,}  \\
&&\r_{l,\GG}(\qq) = \r_{nm\kk,\GG}(\qq) \text{.} 
\eea
The expression for the $KS$ response function is then
\begin{equation}
\xt^{KS}_{\GG\GG'}(\qq,\w)
 =\sum_{l\in\tilde{\Psi}_\qq}  \left|\r_{l,\GG}(\qq)\right|^2 \ft_{l,\qq} \, \Lt_{ll}^{KS}(\qq,\w)  
\label{eq:chi_KS_EQ}
\text{.}
\end{equation}
The response function has contribution only from the transitions for which $\ft_{l\qq}\neq 0$,
corresponding to a pair composed by an occupied $n\kk$ and an empty level $m\kk+\qq$ 
in the ground state. We label this group of transitions as $\tilde{\Psi}_\qq$.
Thus at equilibrium we have the condition $\ft_{l\qq}\neq 0$
if $\{l \in \tilde{\Psi}_\qq\}$.
In Eq.~\ref{eq:chi_KS_EQ} we consider 50 bands, 4 occupied plus 46 empty bands, which
are used to compute $\tilde{W}_{\GG\GG'}(\qq,\w)$.

\begin{figure}[t]
\includegraphics[width=0.45\textwidth]{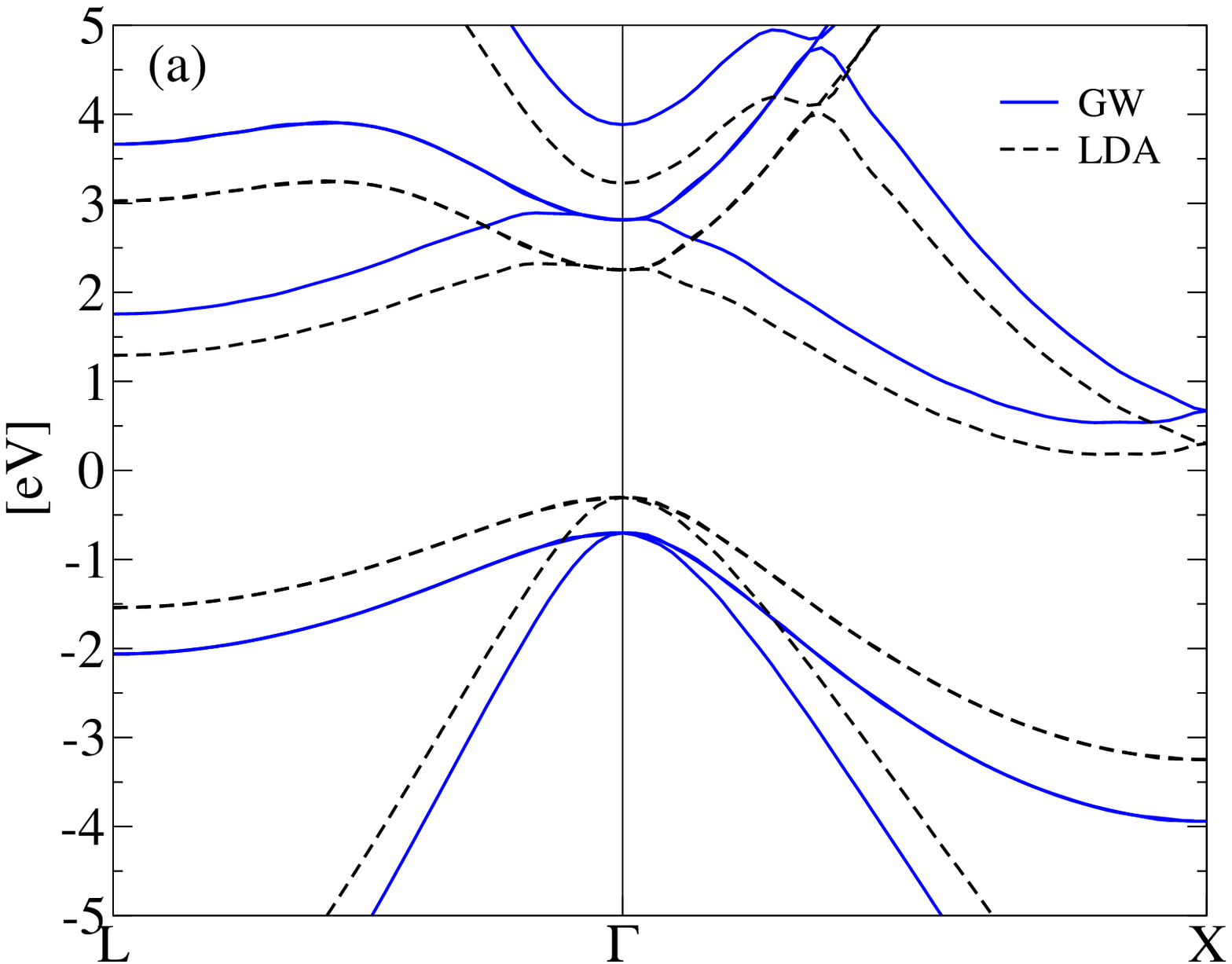}
\includegraphics[width=0.45\textwidth]{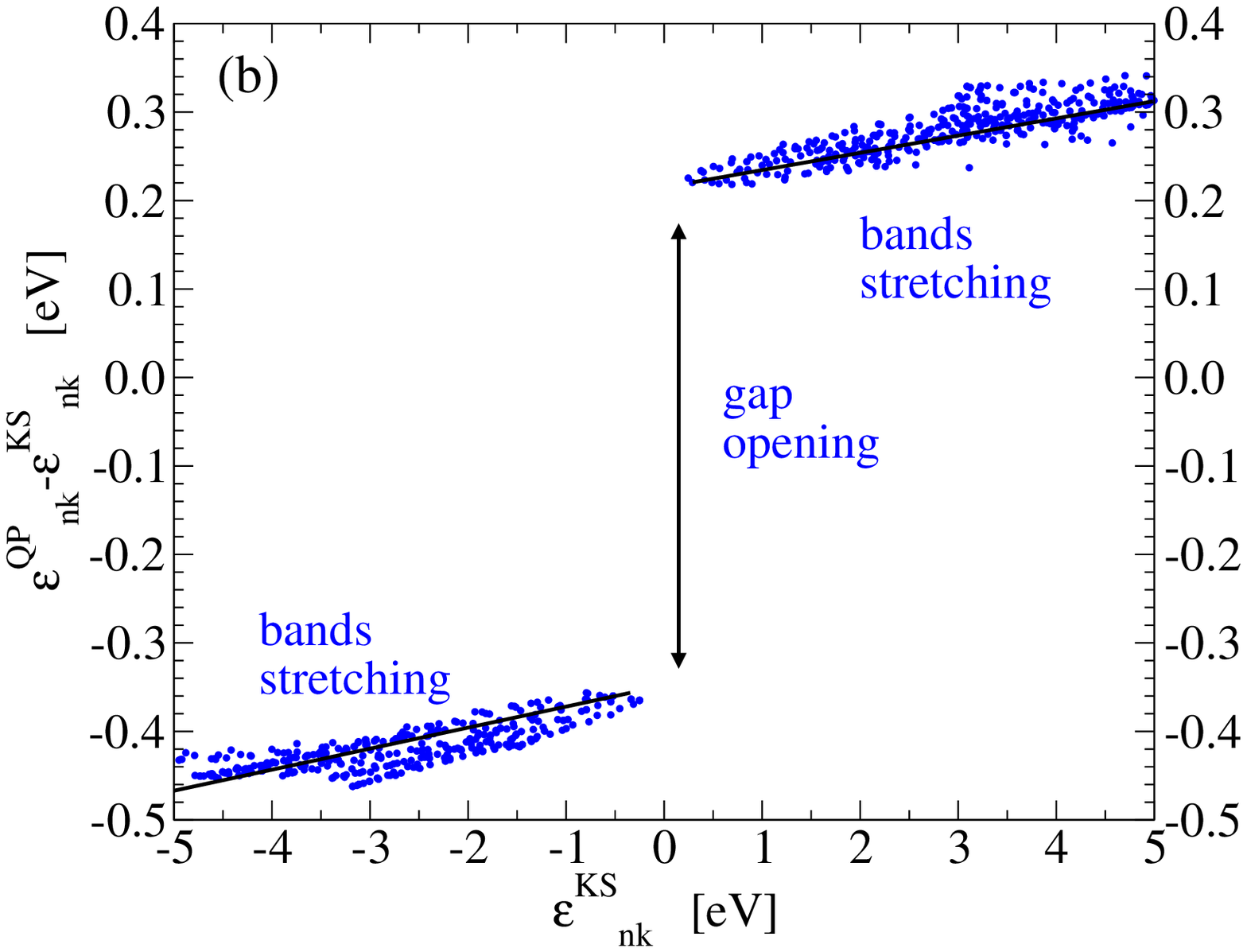}
\caption{(color online) The quasi--particles band structure compared with the Kohn--Sham band structure
         in bulk silicon (panel a) and the difference between the quasi--particles energies
         and Kohn--Sham energies as a function of the Kohn--Sham energies (panel b).}
\label{fig:QP_band_structure}
\end{figure}

In Fig.~\ref{fig:QP_band_structure}.$a$ the $KS$ band structure of silicon
is compared with the QP one. Both band structures are obtained with an interpolation
from the regular $16x16x16$ grid to the $L$-$\G$-$X$ high symmetry lines. 
In Fig.~\ref{fig:QP_band_structure}.$b$ instead
the difference $\e_{n\kk}-\e^{KS}_{n\kk}$ is represented. In both panels
we can appreciate the two principal effects induced by the QP corrections
in semiconductors,
i.e. (i) the opening of the $KS$ band gap and (ii) a stretching of the bands.
In particular the indirect gap is lifted from $0.5\ eV$ to $1.16\ eV$ while the direct
gap from $2.57\ eV$ to $3.43\ eV$. The experimental values are $1.17\ eV$ and
about $3.4\ eV$ respectively~\cite{Hekkwege1982}. Those values are in excellent agreement with the
results of previous calculations~\cite{Rohlfing1993}.

\subsection{Optical properties and the electron--hole interaction}
\label{sec:eq_props:optics}
Excitonic effects are a characteristic feature of the absorption spectra of semiconductors. A well estabilished 
theoretical scheme suitable to describe these effects is the BSE~\cite{Onida2002},
a non linear equation for the four point response function, or electron--hole
propagator, $\Lt^{BSE}(\w)$.
Absorption is defined in the $\qq\ra 0$ limit of the response function.
For simplicity we will thus drop the $\qq=0$ in the notation for energies,
occupation factors and transitions space by defining
$\ft_l\equiv \ft_{l,0}$, $\et_l\equiv \et_{l,0}$ and
$\tilde{\Psi}\equiv\tilde{\Psi}_{\zero}$.

We start by rewriting the BSE in a form where the occupation factors appear explicitly:
\begin{multline}
\Lt^{BSE}_{ll'}(\w)=\Lt^{QP}_{ll'}(\w)+ \\
  \Lt^{QP}_{ls}(\w)\ \sqrt{\ft_s}\Big[v_{ss'}-\tilde{W}_{ss'}\Big]\sqrt{\ft_{s'}}\ \Lt^{BSE}_{s'l'}(\w)
\text{.}
\label{eq:BSE_Dyson}
\end{multline}
In Eq.~\ref{eq:BSE_Dyson} $\Lt^{QP}$ is identical to $\Lt^{KS}$ but for the replacement
$\et_{n\kk}^{KS}\ra\et_{n\kk}$. A summation over the $ss'$ indexes is implicit.
Moreover we have defined 
\bea
\tilde{W}_{ss'}      &\equiv& \sum_{GG'} \r_{nn'\kk,\GG}(\qq) \tilde{W}_{GG'}(\qq,0) \r_{m m'\kk,\GG'}(\qq), \\
v_{ss'}  &\equiv& \sum_{G\neq 0} \r_{nm \kk,\GG}(\qq)\ \ v_G(\qq) \ \ \r_{n'm'\kk,\GG}(\qq).
\eea
${v_\GG(\qq)=4\pi/|\qq+\GG|^2}$ is the bare Coulomb interaction, whose
$G=0$ component is not included.
$\tilde{W}_{\GG\GG'}(\qq,\w=0)$ is the static approximation to the screened
interaction. It is computed with the same parameters used for electron--electron
interaction entering the $GW$ self--energy. Here it describes the direct interaction
between the electron and the hole.

The equation can be re-cast in the form of an excitonic hamiltonian as
\be
\Ht_{ll'}=\et_l - \sqrt{\ft_l} \( v_{ll'}-\tilde{W}^s_{ll'} \) \sqrt{\ft_{l'}},
\label{eq:BSE_exc}
\ee
where we considered, for the index ${l=\{nm\kk\}}$ again a ${16x16x16}$ $\kk$--grid sampling of the
Brillouin zone, 3 occupied states in conduction and 3 unoccupied states in valence band.
The symmetrization via $\sqrt{\ft_l}$, already reported in the literature~\cite{Schleife2011},
ensures that the matrix remains pseudo--Hermitian~\cite{Gruning2011,note_pseudoherm}
also in presence of fractional occupations.
$\Ht$ can then be diagonalized, moving from the electron--hole space
$l$ to the excitonic space $\l$.
The rotation is defined by the eigenvectors $A^{\l,l}$:
\be
\sum_{l'} \Ht_{ll'}\, \At_{\l,l'}=\Et_{\l}\,\At_{\l,l}
\text{.}
\ee
The resulting excitonic propagator $\Lt^{BSE}$
in the excitonic space has the form
\be
\Lt_{\l\l'}^{BSE}(\w)= \frac{S^{-1}_{\l\l'}}{\w-\Et_\l+i\h}
\label{eq:L_BSE}
\text{.}
\ee
$S_{\l\l'}=\sum_{l} \At^*_{\l,l}\At_{\l',l}$ is different from
the identity because the matrix is not Hermitian.

From $\Lt$ the macroscopic dielectric function $\ve$ can be constructed
introducing the macroscopic residuals:
\be
\Rt_{l}= \lim_{\qq\ra 0} \Big( \frac{\r_{l,0}(\qq)}{q} \Big) \sqrt{\ft_l} \text{.} \\
\label{eq:R_macro}
\ee
If we use the independent--particles (IP) response function $\ve$ reads
\be
\vet^{IP}(\w)=1-\sum_{l} \Rt^*_{l} \Lt_{ll}^{IP}(\w) \Rt_{l}
\text{.}
\label{eq:IP-RPA_eps}
\ee
Here $\Lt^{IP}$ can be either $\Lt^{KS}$ or $\Lt^{QP}$.
Eq.\eqref{eq:IP-RPA_eps} is
the $\{\GG=\GG'=\zero\}$ component of Eq.\eqref{eq:eps_micro_RPA}.
We underline here that, thanks to the introduction of the imaginary term $i\h$,
Eq.\eqref{eq:IP-RPA_eps} defines both the real and the imaginary part of $\vet(\w)$,
also respecting the Kramers--Kronig relation.

At the BSE level the macroscopic dielectric function,
and thus the optical properties, are obtained from:
\be
\vet^{BSE}(\w)=1-\sum_{\l\l'} \Rt^*_{\l} \Lt_{\l\l'}^{BSE}(\w) \Rt_{\l'},
\label{eq:BSE_eps}
\ee
where the excitonic residuals are defined as
\be
\Rt_{\l} = \sum_l  \At_{\l,l} \Rt_l 
\label{eq:R_macro_BSE}
\text{.}
\ee
\begin{figure}[t]
\includegraphics[width=0.45\textwidth]{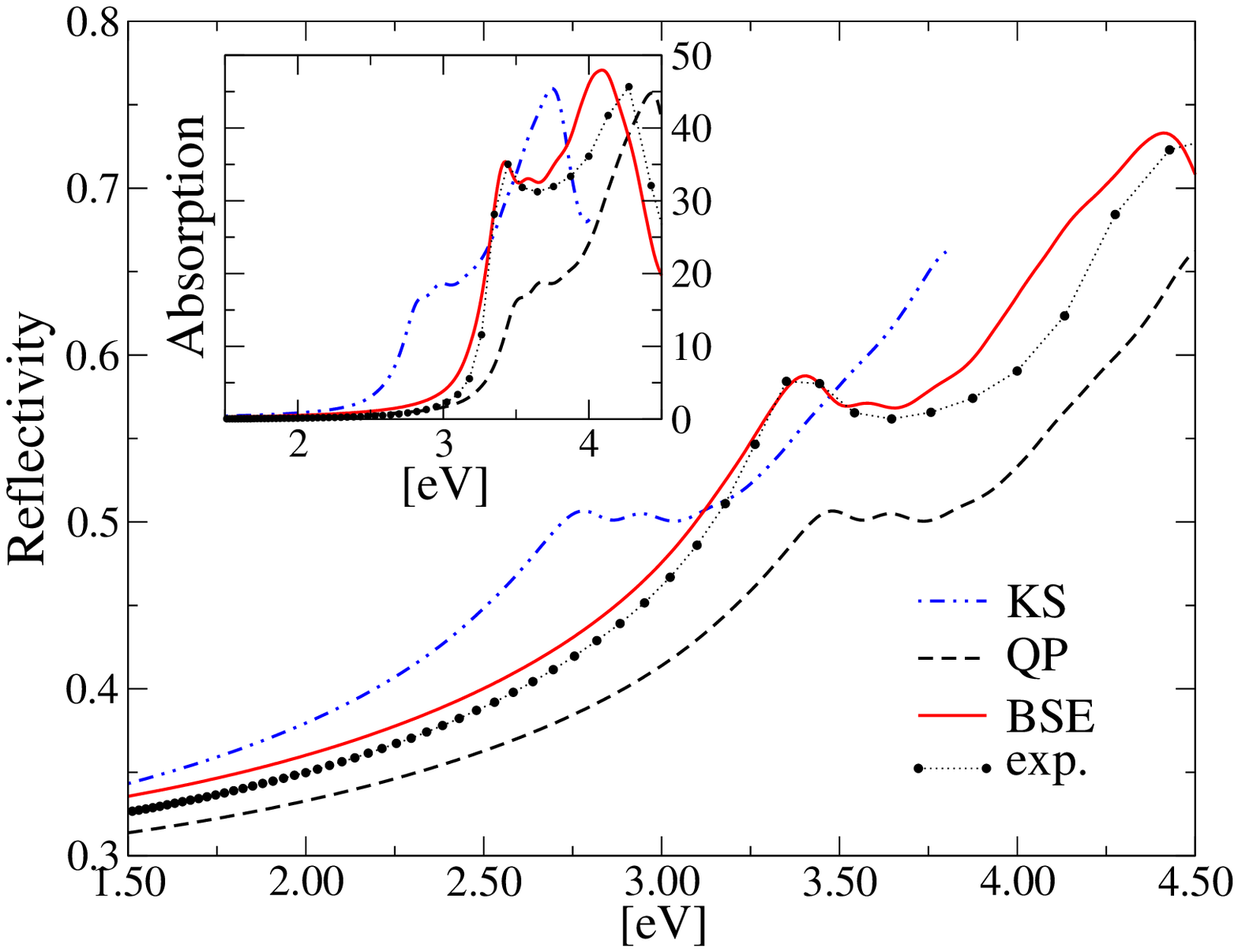}
\caption{(color online) Optical reflectivity of bulk silicon at equilibrium.
         The $KS$ and the $QP$ reflectivity are obtained
         from the Fermi golden rule using Eq.~(\ref{eq:IP-RPA_eps}).
         The BSE spectrum is obtained from Eq.~(\ref{eq:BSE_eps}).
         Experimental data from Ref.~[\onlinecite{Green2005}]. The optical absorption is also
         shown in the inset.}
\label{fig:Optical_properties}
\end{figure}
Fig.~\ref{fig:Optical_properties} reports
the reflectivity at normal incidence defined as
\be
\mathcal{\Rt}(\w)=\left| \frac{\sqrt{\vet(\w)}-1}{\sqrt{\vet(\w)}+1} \right|^2
\label{eq:Refl_def}
\text{,}
\ee
together with the absorption,
i.e. ${\Im[\ve(\w)]}$ (see inset of Fig.~\ref{fig:Optical_properties}).
The results from eq.~(\ref{eq:IP-RPA_eps})
(using either $L_{\l\l'}^{KS}$ or  $\Lt_{\l\l'}^{QP}$)
are compared with the results from eq.~(\ref{eq:BSE_eps})
and with experimental data.
The comparison between the $KS$, the QP and the BSE reflectivity
underlines the twofold role played by the screened interaction.
On one hand the screened electron--electron interaction, included in
the self--energy, opens the electronic band gap, shifting the main peak in the reflectivity
from about $2.8\ eV$ to about $3.5\ eV$. This is the effect of the QP
corrections which we have already discussed in the previous section.
On the other hand the screened electron--hole interaction, which is described
by the kernel of the BSE matrix, closes the optical gap back to about $3.4\ eV$.
The BSE results are in very good agreement with experiments, for what concerns
reflectivity and absorption.

The electronic gap opening can only be described via the long range
electron--electron interaction $W$. It cannot be captured by the LDA $V^{xc}$,
which, instead, is local in space.
Similarly the red shift of the optical gap
and the corresponding rise of the reflectivity intensity from the $QP$ to
the $BSE$ spectrum can be captured only via the long range electron--hole
interaction.
It has indeed been extensively discussed in the literature~\cite{Onida2002,Gonze1995,Varsano2008,Reining2002}
that a kernel local in space, such as the adiabatic local--density approximation,
is not able to capture excitonic effects.

In the NEQ section we will show how the modification  of both the
electron--electron  and of the  electron--hole interactions will affect
the QP states and the BSE poles. We will thus
introduce the concept of optical gap renormalization that,
as a generalization of the well known concept of band gap renormalization,
will be essential to correctly describe the experimental TR signal.

\section{Non--equilibrium optical properties} 
\label{sec:trabs}
In this section we describe the different aspects of the NEQ simulations and the experimental setup.
The experimental data is presented in Section~\ref{subsec:TrAbs_exp}
where the transient reflectivity\,(TR) is measured as a function of the probe frequency $\w$
and of the pump--probe delay $\t$.

In Section~\ref{subsec:neq_dyn} we model the interaction of the system with the pump pulse,
propagating the equation of motion for the $G_{nm\kk}^<(\t,\t)$,
i.e. the density matrix of the system projected onto the $KS$ wave--functions.

In Sections~\ref{subsec:TrAbs_theo_residuals}-\ref{subsec:TrAbs_theo}
the interaction with the probe will be described within the linear response regime.
Since the system is now out of equilibrium, this corresponds to compute the NEQ
dielectric function,  $\ve(\w,\t)$.

As a direct consequence of the definition and calculation of $\ve(\w,\t)$
it will be possible, in  Section~\ref{subsec:TrAbs_theo_residuals}, to introduce
the formal definition of bleaching, photo--induced absorption and  stimulated emission,
beyond the IP approximation.
In Section~\ref{subsec:TrAbs_theo_optical_gap} the different components of the 
optical gap renormalization induced by the pump pulse will be described.
Finally in Section~\ref{subsec:TrAbs_theo} these effects will be considered together
to describe the TR experimental signal.

\subsection{Experimental transient reflectivity in bulk silicon} \label{subsec:TrAbs_exp}

\begin{figure}[t]
\includegraphics[width=0.45\textwidth]{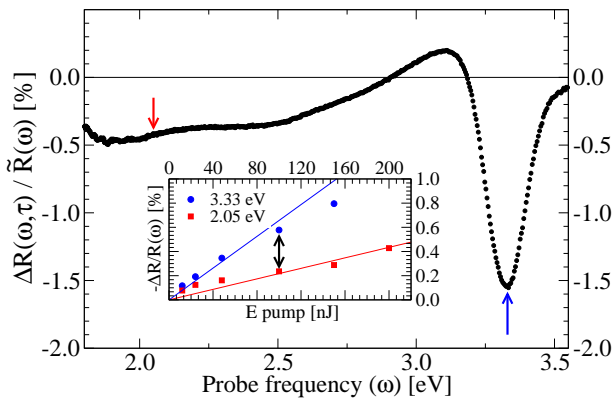}
\caption{(color online) Experimental Transient reflectivity spectrum in bulk silicon at a time
          delay $\t=200\ fs$ between the pump and the probe pulses measured
          at the pump energy of $200\ nJ$, which corresponds to a pump fluence of $1.13\ mJ/cm^2$.
          In the inset the linear dependence of the signal
          as a function of the pump fluence is shown for probe energies of
          $\w=3.33\ eV$ and $\w=2.05\ eV$. These two energies are indicated by the two arrows.}
\label{fig:Transient_reflect_exp}
\end{figure}

One basic technique for time-resolved optical spectroscopy is pump--probe, in which 
a first ultrashort laser pulse (the pump) excites the system.
The whole ensemble of electrons and atoms is thus excited from the ground and unperturbed state
to a time--dependent and complex configuration. At
this point all physical properties (including the change in the optical absorption and reflection)
are dictated by the dynamics of this ensemble, the evolution of which are monitored by a
delayed perturbative laser pulse\,(the probe).
This is used to measure a wealth of time--dependent observables~\cite{Morrison2014,Smallwood2012} of
which transient reflection (TR) is one of the most relevant.

The temporal resolution of the experiment is determined by the duration of
pump and probe pulses. In addition, the need to excite
a system on resonance and probe optical transitions occurring at
different frequencies requires tunability of the pump
pulse, whereas the simultaneous measurement of transmission
changes at many different wavelengths calls for a broadband probe.

The pump--probe setup 
is powered by a regeneratively amplified Ti:sapphire laser system (Quantronix model Integra-C).
The laser source delivers light pulses at $\approx 1.55\ eV$ ($800\ nm$) with duration of $120\ fs$
and at a  repetition rate of $1 kHz$. A fraction of the source is frequency doubled to provide pump
pulses at $\approx 3.1\ eV$ ($400\ nm$) with duration of $100\ fs$ and a total energy which ranges
between $12$ and $200\ nJ$. The pump pulse frequency is close to the optical gap of silicon,
and the thermal broadening of the absorption spectra (see Fig.~\ref{fig:Optical_properties})
allows a direct injection of the photo--excited electronic carriers from the top
of the valence band to the conduction band, around the center of the Brillouin zone ($\Gamma$ point). Another replica of the fundamental source beam is focused
on a $2\ mm$-thick $CaF_2$ crystal, where self-phase modulation generates an ultrabroad white light continuum with spectrum ranging between $1.7\ eV$ and $3.5\
eV$, acting as a probe. The pump and probe pulses are delayed by a high precision delay line and collinearly focused on the sample. The pump laser spot diameter
is estimated to be $\sim$ 150 $\mu m$ which corresponds to a fluence of $\approx 1.13 mJ/cm^2$ at an energy of $200\ nJ$. The probe spot diameter is about half
that of the pump. The TR signal, \emph{i.e.} the relative variation of the reflectivity from the equilibrium,
is obtained by  measuring, at single-shot rate,
the reflected probe spectra with ($\mathcal{R}(\w,\t)$) and without ($\mathcal{\Rt}(\w)$) the pump excitation.
($\mathcal{R}(\w,\t)$) is collected at several pump--probe delays $\t$
\cite{Polli2007}, defined as the peak to peak distance between the pump and the probe laser pulse.
The TR signal is then defined as 
\be
\frac{\D \mathcal{R}\(\w,\t\)}{\mathcal{\Rt}\(\w\)} = \frac{\mathcal{R}(\w,\t)-\mathcal{\Rt}(\w)}{\mathcal{\Rt}(\w)} \text{.}
\label{eq:Tr_Refl_def}
\ee
The ultrabroad spectrum of the white light continuum and the unique single-shot spectrometer
allows to simultaneously probe the evolution of the system over an unprecedentedly broad frequency range. 

Figure \ref{fig:Transient_reflect_exp} shows the measured $\D \mathcal{R}/\mathcal{\Rt}$ spectrum
at a delay time of $200\ fs$, \emph{i.e.} when the relative variation of the signal is maximum.

The optical response has been explored in a wide range of pump fluencies
showing a linear behavior within the entire range of probe photon energies as reported in the inset of
Fig.~\ref{fig:Transient_reflect_exp}.
The $\D \mathcal{R}/\mathcal{\Rt}$ spectrum is dominated by a pronounced narrow negative peak centered
around $3.3\ eV$. The low energy side of the peak displays interesting features
at energies lower than the optical gap: between $2.9\ eV$ and $3.2\ eV$ the signal becomes
positive while below $2.9\ eV$ down to the near-IR optical region is completely negative.
In the reflection geometry it is not easy to directly associate these features to clear
physical effects, since both the real and the imaginary part of the dielectric function $\ve(\w,\t)$
enters the definition of the reflectivity.
The negative peak could be related to the Pauli blocking of the optical
transition (or bleaching effect), resonant to the gap.
The negative signal, below the optical gap, could be instead assigned to a variation of the
$\Re[\ve(\w,\t)]$. Indeed $\Im[\ve(\w,\t)]$ below the optical gap, and its variation,
could be related to transient state filling effects involving phonon or defect mediated
optical transitions and is expected to be some order of magnitude smaller.

\subsection{Theory: pump pulse and carriers generation} 
\label{subsec:neq_dyn}
In order to properly simulate transient experiments the effect of both the pump and the 
probe pulses must be included.
The approach we follow here is to describe the effect of the primary pump field 
by solving the Kadanoff-Baym equation\,(KBE),
i.e. the equation of motion for the NEQ Green function $G_{nm\kk}^<(t,t')$. The knowledge of the $G^<$ function provides access to a wealth of theoretical observables and it is also used, indeed, to calculate the probe field response.
As the probe is much weaker than the pump we can safely work in the linear--response regime and
compute the TR signal defining a response function functional of the $G^<$.    
Under specific conditions, the TR becomes a functional of the sole non--equilibrium occupations at a 
specific time and the NEQ response can be obtained solving the NEQ--BSE~\cite{Perfetto2015a},
whose solution gives direct access to the transient spectrum measured by the probe pulse.

These approximations are:
(i) the generalized Kadanoff--Baym ansatz~\cite{HartmutHaug2008,Bonitz1998,Stefanucci2013,Kadanoff1962},
(ii) a Markovian approximation for the collisions integral, known as complete collision approximation~\cite{HartmutHaug2008,Marini2013},
(iii) an adiabatic approximation for the functional dependence on the density matrix~\cite{Perfetto2015a} and last,
(iv) the assumption that the excited carriers form a QP electron--hole gas.

The approximations (i)--(iii) have been already discussed in the literature.
The adiabatic approximation, in particular, is the core result of Ref.~\onlinecite{Perfetto2015a}.
The idea is that the probe laser pulse is used to take a picture of the system. If the system
does not evolve during the interaction with the pulse, it can be considered in a
quasi--stationary state at the instant $\t$, described by $G^<(\t,\t)$. This enables to take
$G^<(\t,\t)$ outside the Fourier transform against the probing time and thus define a NEQ dielectric
function $\e(\w,\t)$ which we compute in the next section. In case the system evolves in such period
the experimental picture would result blurred compared to the theoretical one, similarly to a 
picture taken onto an object moving faster than the exposure time.

Under these three assumptions,
we can consider the time diagonal $G^<(\t,\t)$, i.e. the density matrix,
and project it on the $KS$ (or the QP) space, i.e.
$G^<_l(\t)\equiv G^{<}_{nm\kk}(\t)$.
Its equation of motion reads:
\begin{multline}
\partial_t G_l^<=\et_l\ G_l^<+ \\ [\D\S^{Hxc},G^<]_l+[U^{pump},G^<]_l+S_l[G^<]
\text{.}
\label{eq:KBE}
\end{multline}
The ${l=\{nm\kk\}}$ index run on a $16x16x16$ $\kk$--grid of the
Brillouin zone, with 3 valence bands and 3 conduction bands. Since the KBE is solved in the
$KS$ space these are the only parameters needed.

$\et_l$ are the QP energies differences at equilibrium (see Eq.\eqref{eq:delta_E_f}).

$\D\S^{Hxc}\equiv \S^{Hxc}-\tilde{\S}^{Hxc}$, is the variation of the Hartree potential and the
COulomb Hole plus Screened--EXchange (COHSEX) self--energy ($\S^{Hxc}$).
$S_l[G^<]$ describes instead the relaxation and dissipation
processes~\cite{Marini2013}.
In the COHSEX approximation the variation of the $GW$ self--energy is treated within the
static approximation (i.e. the screened interaction is assumed static and constant during the time propagation). This constrain is essential to
correctly describe the coupling with the pump pulse~\cite{Attaccalite2011}.
Numerically, the self--energy and the Hartree potential
are computed using wave--functions defined
on a $4x4x4$ and then interpolated on a bigger $16x16x16$ $\kk$--grid by
using a nearest--neighbor interpolation scheme.

$U^{pump}(t)=-V \efield(t)\cdot\PP$ is the term which describes the pump pulse
within the dipole approximation in the length gauge.
$\PP$ is the polarization per unit volume and
$\efield(t)$ the pump pulse electric field.

In order to mimic the experimental conditions, we used a pump pulse with Gaussian envelope profile and 
a carrier wave oscillating at $750\ THz$ frequency,
corresponding to $3.1\ eV$ photon energy.
The full width at half maximum of the envelope is $100\ fs$.
The pump peak intensities we use range from $10^7$ to $10^{11}$ $W/cm^2$.
The final comparison with the experimental TR
is done for an intensity of
$10^9\,W/cm^2$, which corresponds to a fluence of $0.0886\ mJ/cm^2$.

In our equation of motion we do not directly include the induced macroscopic electric field.
This corresponds, for example, to
the transverse approach used in Ref.~\onlinecite{Yabana2012}
Such approach implies that the simulation field corresponds to the total
field inside the medium. It can be
related to the incident pulse electric field via the relation~\cite{Yabana2012}
\be
\efield^{inc}(\w)=\frac{1+\sqrt{\vet(\w)}}{2}\efield(\w)
\text{.}
\label{eq:induced_E}
\ee
Eq.\eqref{eq:induced_E} is obtained by subtracting from the incident field its reflected part.
As $Re[\ve(\w=3.1)]\approx 33$ we have that, neglecting the imaginary part $Im[\ve]$ which is smaller
in Eq.\eqref{eq:induced_E}, $\sqrt{\vet(\w)}\sim\,11.4$.
It follows that our simulations should be compared with an
experimental pulses of $0.0886\times11.4\approx 1\ mJ/cm^2$.
This value turns out to be very close
to the experimental fluence of $1.13\ mJ/cm^2$ used in this work.

\begin{figure}[t]
\includegraphics[width=0.45\textwidth]{Fig4a.eps}
\includegraphics[width=0.45\textwidth]{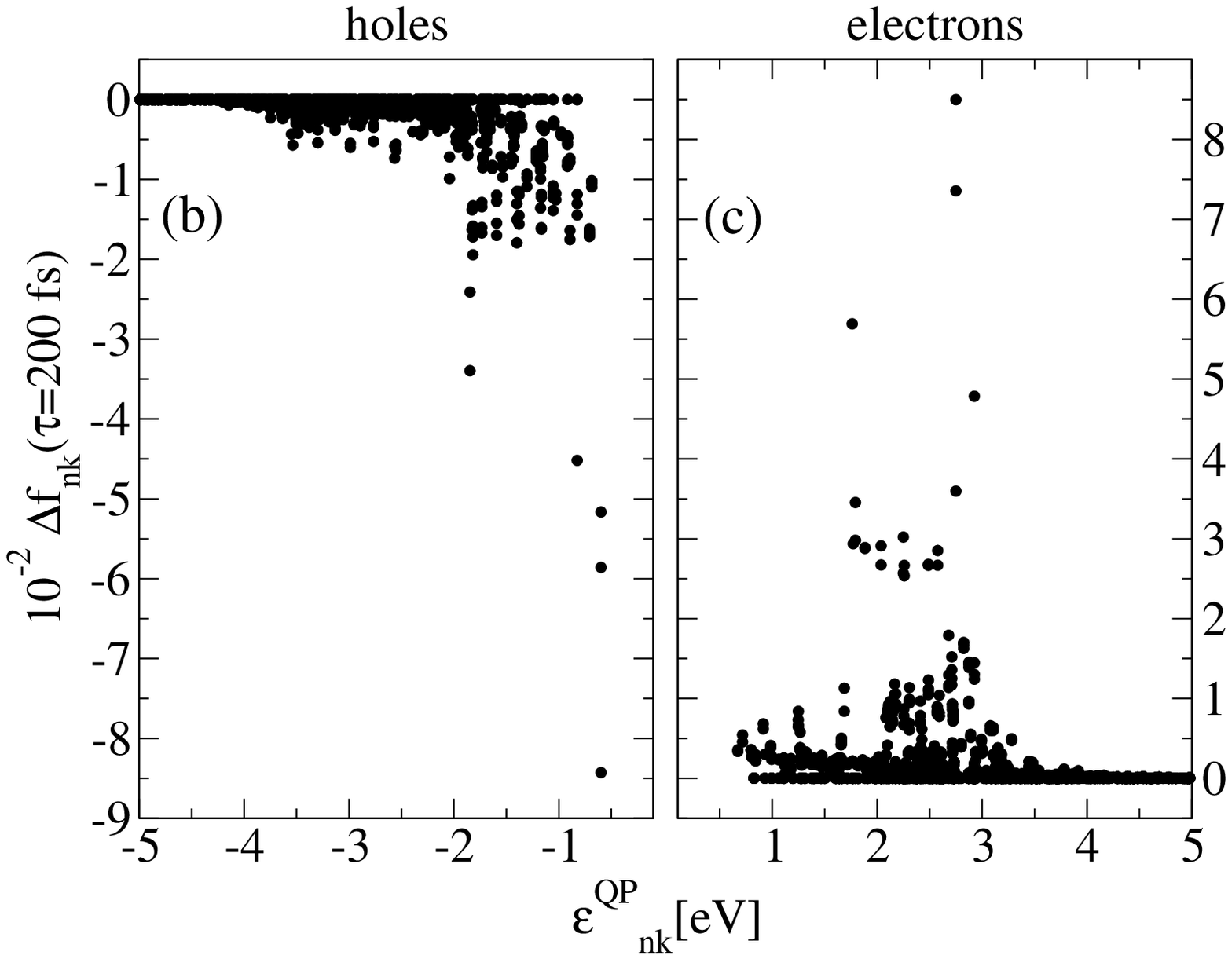}
\caption{(color online)
         The total number of carriers (panel $a$) as a function of the laser pulse
         intensity (red circles) and the maximum change in the electronic
         occupations (blue squares). In the inset the total number of
         carriers as a function of time and the pump pulse are shown for the pump
         peak intensity of $10^9$ $W/cm^2$, which corresponds to
         an estimated pump fluence for the incident pulse of $\approx 1\ mJ/cm^2$.
         The holes (panel $b$) and electrons distributions (panel $c$)
         at $\t=200\ fs$, for the same pump peak intensity is also shown. 
         }
\label{fig:NEQ_carriers}
\end{figure}

The last approximation used is the electron--hole QP gas approximation which assumes that the excited state carriers form a gas of hot electrons. 
This ansatz is well justified because, in our simulations, the carriers densities are above
$\approx 10^{16}$ $el/cm^3$, which is the critical Mott density, $\r_M$ of silicon~\cite{Norris1982}.
It is indeed well known that for densities below $\r_M$ the carriers are trapped in excitonic
states and do not behave anymore as delocalized Bloch electrons.
Within this approximation the off--diagonal elements
of $G_l^<(\t)$, i.e. for $n\neq m$, which describe the polarization,
can be assumed to rapidly decay  due to dephasing effects.
Therefore, when the probe response is measured, only the diagonal elements survive and define the NEQ occupations
\be
f_{n\kk}(\t)=\Im[G^{<}_{nn\kk}(\t)]
\text{.}
\label{eq:occ_neq}
\ee
From Eq.\eqref{eq:occ_neq} it also follows that ${f_{l,\qq}(\t)=f_{m\kk+\qq}(\t)-f_{n\kk}(\t)}$.

The number of carriers created by the pump pulse is given by the difference
with the equilibrium population, i.e. $f_{n\kk}(\t)-\ft_{n\kk}$.
In the present case the  carriers densities excited by the 
experimental pump 
range from
$10^{17}$ to $10^{22}\ el/cm^3$.
More precisely, at  a pump intensity of $10^9$ $W/cm^2$ the
carriers density is $8\cdot 10^{19}$ $el/cm^3$
(see Fig.~\ref{fig:NEQ_carriers}.$a$ and its inset). 

Fig.~\ref{fig:NEQ_carriers}.$a$ demonstrates the linear dependence
of the NEQ carriers excited by the pump pulse as a function of its intensity.
This ensures a linear dependence of the TR signal characteristic of the
low density regime.
As shown in the figure, only above $10^9$ $W/cm^2$ the
number of carriers deviates from the linear regime.
The reason is that at these very high pump intensities the
carriers approach the maximum occupation ($f_{n\kk}\sim\,0.5$) which can be induced by the pump
pulse. At this high laser intensities the effect of the pump is not only
to excite carriers from the valence to the conduction band, but also
to stimulate the emission back from the conduction to the valence.

In the following we will consider a pump--probe delay of $\t=200\ fs$, corresponding to the experimental condition.
If we neglect the carriers relaxation (this means we set $S_l=0$ in eq.~\ref{eq:KBE})
the electrons and holes NEQ populations are showed in 
Fig.~\ref{fig:NEQ_carriers}.$b$-$c$.

These computed NEQ occupations will be used, in the next sections,
to highlight the renormalization of the
different components of the dielectric function.
More precisely their effect will be described in two contributions.
The first is the variation of the \emph{optical residuals} defined by Eqs.~\eqref{eq:R_macro} and
\eqref{eq:R_macro_BSE} which will lead to a formal definition of the concepts of bleaching,
photo--induced absorption and stimulated emission (Sec.~\ref{subsec:TrAbs_theo_residuals}).
The second is, instead, the variation of the many--body self--energy and of the BSE kernel introduced in
Eqs.~\eqref{eq:GW_eq} and \eqref{eq:BSE_exc} which will lead to the definition of the
{\em optical--gap renormalization} (Sec.~\ref{subsec:TrAbs_theo_optical_gap}).
Once these effects have been defined in terms of the dielectric function, then they will enter
the computation of the reflectivity and thus gives the theoretical TR of silicon.

\subsection{Theory: probe pulse and residuals renormalization}
\label{subsec:TrAbs_theo_residuals}
As we have already observed in sec.~\ref{subsec:TrAbs_exp} the TR signal is not easily
associated with physical processes since $\mathcal{R}(\w,\t)$ depends on both $\Im[\ve(\w,\t)]$
and $\Re[\ve(\w,\t)]$. Thus in this section we first discuss the variations of
the dielectric function $\ve(\w,\t)$. The TR signal is then derived using
Eqs.~\eqref{eq:Refl_def} and \eqref{eq:Tr_Refl_def}.

The NEQ dielectric function can be obtained from the IP response function
replacing the equilibrium residuals with the NEQ ones in Eq.\eqref{eq:IP-RPA_eps}:
\be
\ve^{IP}(\w,\t)
 =1-\sum_{l\in\Psi^\t} 
  R^*_{l}(\t) \Lt_{ll}^{IP}(\w) R_{l}(\t)
\label{eq:IP-RPA_eps_var}
\text{.}
\ee
As mentioned earlier $\Lt^{IP}$ can be either the $\Lt^{KS}$ equilibrium response function
or the $\Lt^{QP}$ one defined in terms
of the equilibrium QP energies.
The NEQ residuals are defined, within the adiabatic approximation~\cite{Perfetto2015a} as:
\be
R_{l}(\t) = \lim_{\qq\ra 0} \Big( \frac{\r_{l,0}(\qq)}{q} \Big) \sqrt{f_l(\t)} \text{.} 
\ee
Eq.~(\ref{eq:IP-RPA_eps_var}) 
can be naturally split in two terms according the the indexes in the summation.

{\em Equilibrium part}: the transitions corresponding to indexes $l$ that belong to the 
equilibrium group ($\{ l \in \tilde{\Psi} \}$) define the \emph{bleaching effect}.
In this case the reduced number of electrons in the occupied
states (and holes in the unoccupied states) decrease the absorption
signal~\footnote{$Im[\ve(\w)]$ describes in general both absorption and stimulated emission.
       However, considering the transitions possible from the ground state, i.e. at equilibrium,
       the associated absorption is maximum while there is no stimulated emission. For this reason
       in the literature $Im[\ve(\w)]$ is commonly identified with the absorption only.
       Out of equilibrium, instead, absorption decrease while stimulated--emission
       becomes different from zero. Thus bleaching includes, rigorously speaking,
       the variation of these two effects and not a reduction of the absorption.}.

{\em Photo--induced part}: when the index $l$
belong to new transitions induced  not possible at equilibrium
($\{ l \notin \tilde{\Psi} \}$). 
It describes \emph{photo--induced absorption} when the energy
of the final level is greater than the energy of the starting level
(${\e_{n\kk}>\e_{m\kk}}$), and 
the associated \emph{stimulated emission} vice--versa
(${\e_{n\kk}<\e_{m\kk}}$).

A simple and direct extension to the excitonic case can be derived
when the variations of the occupations factors
are small,  $|f_{n\kk}(\t)-\ft_{n\kk}|<0.01$.
In this case we can assume that the QP energies,
the excitonic poles $\Et_\l$ and the wave--functions $\At^\l_l$ do not
change following the pumping process. We can, therefore, define the space of
equilibrium excitonic transitions $\{\l\}$
(i.e. the space of the excitonic indexes obtained diagonalizing the equilibrium excitonic matrix)
as $\l\in\tilde{\Phi}$ and
\begin{multline}
\ve^{BSE}(\w,\t)
  \approx 1-\sum_{\l\l'\in\tilde{\Phi}} R^*_{\l}(\t) \Lt_{\l\l'}^{BSE}(\w) R_{\l'}(\t)  \\
    -\sum_{l \notin \tilde{\Psi}} R^*_{l}(\t) \Lt_{ll}^{QP}(\w) R_{l}(\t)
\text{,}
\label{eq:BSE_eps_var_RES}
\end{multline}
with the NEQ residuals:
\be
R_{\l}(\t)=\sum_{l \in \tilde{\Psi}} \At_{\l,l} R_{l}(\t) \text{ where }\l\in\tilde{\Phi}
\text{.}
\ee
Note that the kernel of the excitonic matrix at the equilibrium are different 
from zero only in the case ${\{ l \in \tilde{\Psi}\}}$, i.e. if ${\ft_l\neq 0}$
(see Eq.~\eqref{eq:BSE_exc}).
Thus only the ``equilibrium part'' of the absorption is written in terms of excitonic
poles while  the ``induced part'' is identical to the QP energy differences.

\begin{figure}[t]
\includegraphics[width=0.45\textwidth]{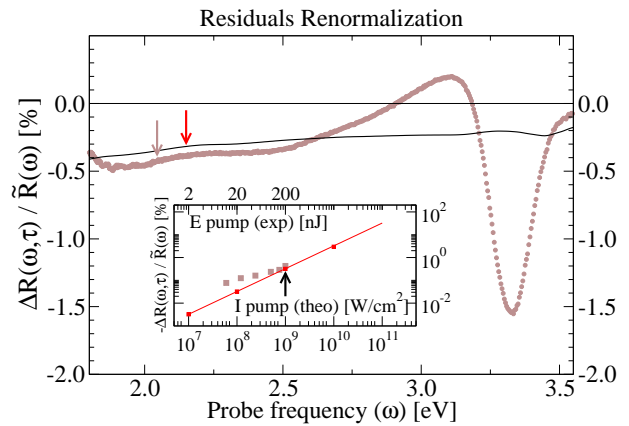} 
\caption{(color online) Calculated\,(line) versus measured\,(dots) transient reflectivity for a
         pump intensity of $10^9 W/cm^2$  and $\t=200\,fs$ assuming the BSE poles remain unchanged.
         In the inset the signal at $2.1\ eV$ is
         shown as a function of the pump peak intensity.
         The calculation\,(line) reveals a perfect linear regime and are compared with the experiment\,(boxes).
         Nevertheless the transient reflectivity  is only qualitatively described and the simulation
         misses the peculiar behaviour of the experimental spectrum.}
\label{fig:Transient_reflect_theo_Ronly}
\end{figure}

The computed TR of silicon 
is shown in Fig.~\ref{fig:Transient_reflect_theo_Ronly}. The main effect
is the bleaching of the first absorption peak, due to a Pauli blocking effect
induced by the NEQ occupations. Thus the variation of the imaginary part of the
dielectric function (not shown) displays a negative peak at approx $3.3\ eV$.
Due to the Kramers--Kronig relations, this corresponds
to a reduction of the real part of $\ve(\w,\t)$. This explains the overall
reduction of the reflectivity with a constant negative TR signal.
This effect well describes the negative TR in the low
probe photon energy region, but fails to capture the pronounced negative peak close to
the optical gap at about $3.35\ eV$ and the associated positive peak.

\subsection{Theory: probe pulse and optical gap renormalization}
\label{subsec:TrAbs_theo_optical_gap}
From  Fig.~\ref{fig:Transient_reflect_theo_Ronly} it is clear that the assumption of
un--perturbed excitonic energies (and wave--functions) is not valid.
For a better description of the TR signal we need to compute
their variation due to the presence of additional NEQ carriers.
Since we consider the excitonic energies 
we also need to consider the variation of the underlying
QP band structure, $\et_{n\kk}$.

We underline that the position of the absorption
peaks corresponds to the poles of the many--body response function which
are uniquely determined by the many--body Hamiltonian and cannot
change, as it has been shown by recent calculations on a model
system with few poles~\cite{Johanna2015}. Thus any variation induced
in the single--particle energy levels of the system by the NEQ occupations
has to be exactly compensated in the calculation of the poles of the response
function.
However when complex systems, such as extended semi--conductors, are considered,
the number of poles increases and the description in terms of all the poles of
the many--body Hamiltonian is not convenient anymore. Such systems are thus described in terms
of QP with effective complex energies which are assumed to capture
the effects of the interaction with the rest of the system.
Accordingly the corresponding response function constructed from the BSE
is described in terms of effective excitonic poles.

The variation of such effective poles is not in contrast with the fact
that the real poles do not change.
Indeed such variation, known in the literature as band--gap renormalization,
have already been measured in different materials~\cite{Hase1}.
However, to get a meaningful picture, 
the renormalization of the QP and excitonic poles must be treated consistently.

We show here that, within NEQ \ai-MBPT, both the renormalization of the QP
energies and of the excitonic poles, is dictated by the simultaneous effect
of the change in the screened interaction and in the carriers
population induced by the pump pulse.
Thus in both cases we use a 
static approximation to the screened electron--electron and electron--hole
interaction.~\footnote{Another option maybe to follow the same approach used to
compute the equilibrium spectrum, i.e. to use a dynamical screening for the QP corrections
and a static screening to describe the electron--hole interaction. We tried to consider
the effect of dynamical corrections, re--normalizing the variation
of the COHSEX energies via the equilibrium Z factors computed within the GW approximation.
while this did not change substantially the position of the negative
peak in the TR, led to a general worsening of the agreement with the experimental spectrum}.
By following the adiabatic approximation~\cite{Perfetto2015a} we introduce the 
NEQ COHSEX self--energy and the NEQ BSE. To this end we first define the NEQ
statically screened interaction, $W^\t$ as obtained by 
inserting the NEQ occupations $f_l(\t)$ in
Eq.~(\ref{eq:chi_KS_EQ}).

The static COHSEX approximation for the  $GW$ self--energy then follows:
\begin{multline}
\S^{\t}_{cohsex}(\xx,\xx')=\\\int d^3\kk\ \[ e^{i\kk(\RR-\RR')} 
               \times \int d^3\qq\ \g^\t_{\kk+\qq}(\rr,\rr')  W_{\qq}^\t(\rr,\rr',0)\] \\
                +\d(\xx-\xx') \int d^3\qq\ e^{i\qq(\RR-\RR')} W_{\qq}^\t(\rr,\rr',0)
\text{,}
\label{eq:NEQ_COHSEX}
\end{multline}
with
\be
\g^\t_{\kk}(\rr,\rr')=2\sum_n f_{n\kk}(\t)\psi^{KS,*}_{n\kk}(\rr)\psi^{KS}_{n\kk}(\rr') .
\label{eq:gamma_KS}
\ee
Thanks to the inclusion of the time--depedent occupations in $W^\t$ also the self--energy,
and thus the QP energies, acquire a $\t$ dependence.
Therefore, quite in general, the variation of the QP corrections can be expressed as
\be
\D\e^\t_l= \langle l | \S^{\t}_{cohsex}-\tilde{\S}_{cohsex} | l \rangle.
\label{eq:NEQ_QP-COHSEX}
\ee
From Eq.\ref{eq:NEQ_QP-COHSEX} we can also define the NEQ BSE, which reduces to the
computation of eigenvalues\,($E^\t_\l$) and eigenvectors\,($A^\t_{\l,l}$) of
the NEQ $\t$--dependent excitonic matrix:
\be
H_{ll'}^{\t}=\et_l+\D\e^\t_l - \sqrt{f_l(\t)} \( v_{ll'}-W^{\t}_{ll'} \) \sqrt{f_{l'}(\t)}.
\label{eq:NEQ-BSE_exc}
\ee
Eq.~\eqref{eq:NEQ-BSE_exc} defines
$L^{BSE}(\w,\t)$ which is the NEQ version of $\Lt(\w)$ and can be written in the basis defined
by the excitonic wave-functions $A^\t_{\l,l}$.

It is essential to note, here, that Eq.\eqref{eq:NEQ-BSE_exc} lives in a transition space larger then in the equilibrium case. This new space,
${l \in \Psi^\t}$, includes also the new trasnsitions created by the pump pulse defined by the condition
$f^\t_l\neq 0$.
Numerically, then, the NEQ BSE corresponds to a larger matrix to diagonalize
than the standard BSE.

In the low density regime, however, we can assume that only the excitonic energies
change and not the BSE wave--functions. Indeed energies changes already at the first order
in the perturbation, while wave--functions only at the second order.
Thus have $A^\t_{\l,l}\approx \At_{\l,l}$ and
the residuals are thus independent from the pump--probe delay $\t$
\be
\sum_{l\in \tilde{\Psi}} \At^\t_{\l,l} \Rt_{l}
 \approx \sum_{l\in \tilde{\Psi}} \At_{\l,l} \Rt_{l}  = \Rt_{\l} 
\text{.}
\ee
The index $l\in \tilde{\Psi}$ since $\Rt_{l}=0$ if 
$l\notin \tilde{\Psi}$ and 
in practice only bleaching effects are included if the
oscillator strengths are assumed to be at equilibrium.

In this limit the NEQ absorption has the form:
\be
\ve^{BSE}(\w,\t)\approx 1-\sum_{\l\l'\in\tilde{\Phi}}  \Rt^*_{\l} L_{\l\l'}^{BSE}(\w,\t) \Rt_{\l'}
\label{eq:BSE_eps_var_MBPT}
\text{.}
\ee
$L(\w,\t)$ depends on $\t$ via the NEQ occupations $f(\t)$, which enter
Eqs.~\ref{eq:gamma_KS} and \ref{eq:NEQ-BSE_exc}, and via the NEQ screened interaction
which enters Eqs.~\ref{eq:NEQ_COHSEX} and \ref{eq:NEQ-BSE_exc}.

\begin{figure}[t]
\includegraphics[width=0.45\textwidth]{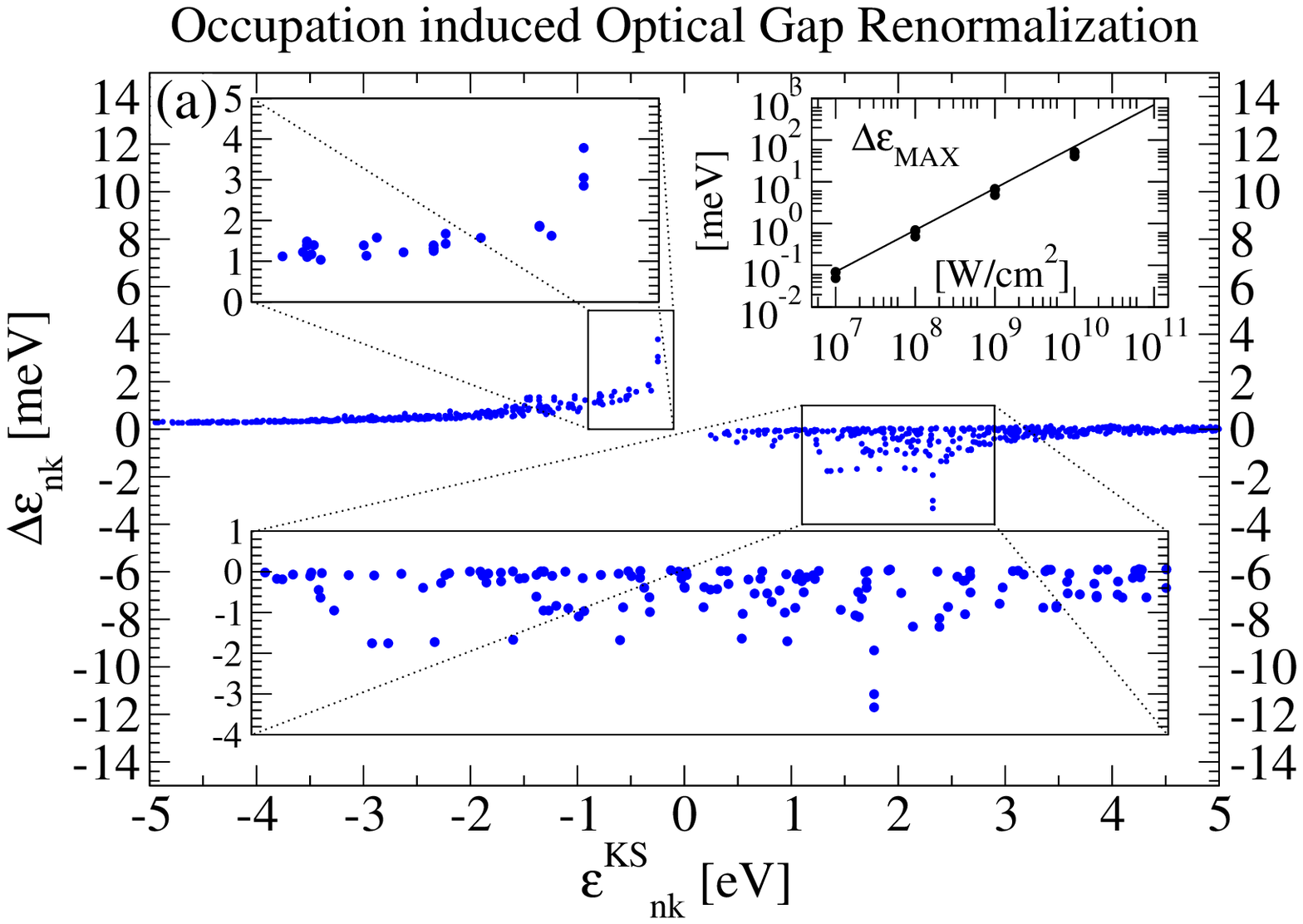}
\includegraphics[width=0.45\textwidth]{Fig6b.eps} 
\caption{(color online) Renormalization of the quasi--particles energies (panel a)
         from the COHSEX self--energy computed with the equilibrium screening.
         Transient reflectivity (panel b) for a
         pump intensity of $10^9\ W/cm^2$ and $\t=200\,fs$; the variation of
         the BSE poles at fixed screening is considered.
         In the insets the maximum change in the energy levels (panel a) and
         the signal at $3.3\ eV$ (panel b) are
         shown as a function of the pump peak intensity. Theoretical results
         are compared with experimental data (brown circles).}
\label{fig:Transient_reflect_theo_Gonly}
\end{figure}

\begin{figure}[t]
\includegraphics[width=0.45\textwidth]{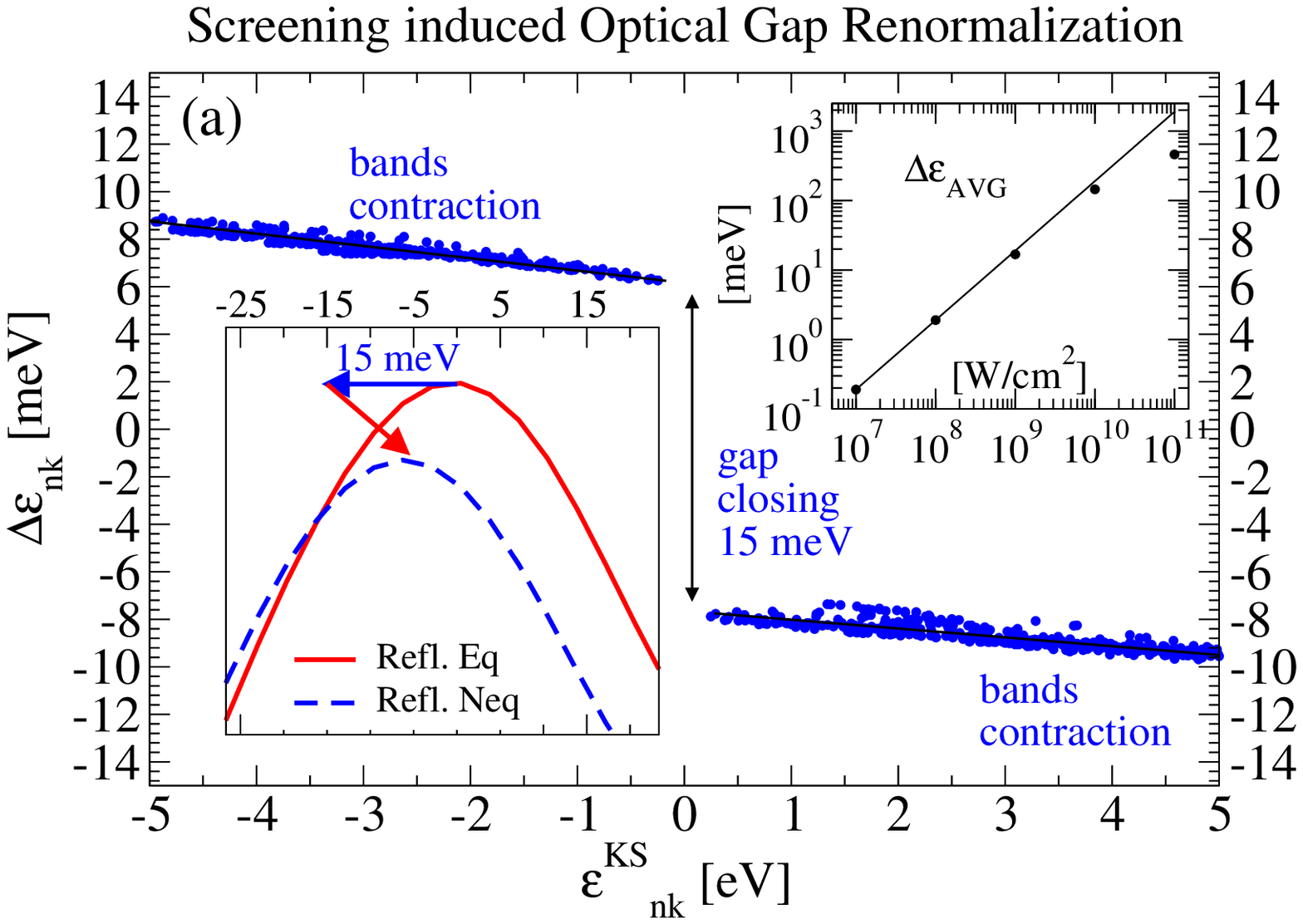}
\includegraphics[width=0.45\textwidth]{Fig7b.eps} 
\caption{(color online) Renormalization of the quasi--particles energies (panel a)
         from the COHSEX self--energy due to the screening of the non--equilibrium
         carriers density. The change induced in the reflectivity near the main peak
         is shown in the inset (blue arrow) together with the effect due to the
         renormalization of the electron--hole interaction (red arrow).
         Transient reflectivity (panel b) for a
         pump intensity of $10^9\ W/cm^2$ and $\t=200\,fs$; the variation of
         the BSE poles due to the screening of the non--equilibrium
         carriers density is considered.
         In the insets the change in the gap (panel a) and
         the signal at $3.45\ eV$ (panel b) are
         shown as a function of the pump peak intensity. Theoretical results
         are compared with experimental data (brown circles).}
\label{fig:Transient_reflect_theo_Xonly}
\end{figure}

\emph{Occupations induced renormalization.}
We first consider the case where we keep the equilibrium
RPA screening, i.e. we use $\tilde{W}$ instead of $W^{\t}$. 
Thus the $\t$ dependence of $L(\w,\t)$ enters only via the NEQ occupations $f(\t)$.
We refer to this as occupations induced renormalization.

In Fig.~\ref{fig:Transient_reflect_theo_Gonly}.$(a)$ the variation of the QP energies
and of the reflectivity are shown under these assumptions.
We notice that the QP corrections follow the 
NEQ carriers distribution and are very small, only few $meV$s.
The resulting TR signal, shown in Fig.~\ref{fig:Transient_reflect_theo_Gonly}.$(b)$,
presents a derivative feature which can be understood as a blue energy shift.
However the signal is very small if compared to the experimental
data and thus the global effect is, on the average, negligible.

\emph{Screening renormalization.}
We then consider the effect on the TR due to the variation of the RPA screening only,
i.e. the $\t$ dependence of $L(\w,\t)$ enters only via $W^\t$
while the equilibrium occupation factors $\ft_l$ are used.

In Fig.~\ref{fig:Transient_reflect_theo_Xonly} the variation of the 
QP energies (panel $a$) and of the reflectivity spectrum  (panel $b$) are shown.
First of all, comparing Fig.~\ref{fig:Transient_reflect_theo_Gonly} with 
Fig.~\ref{fig:Transient_reflect_theo_Xonly} we clearly see that the change in the 
QP energies and in the TR induced by the variation of the screening is much bigger
and this time of the same order of the experimental signal.

Due to the presence of the NEQ carriers the screening increases.
The global effect reverses what we saw in the equilibrium case
(Figs.~\ref{fig:QP_band_structure}-\ref{fig:Optical_properties}).
Thus the electronic gap is closed, the bands dispersion is contracted and 
the reflectivity spectrum is brought ``back'' towards the $KS$ one.
We name this effect as optical gap renormalization.
The optical gap renormalization enters in the same way in the real and in the imaginary part 
of the dielectric function and thus in the reflectivity. 
As shown in the inset of Fig.~\ref{fig:Transient_reflect_theo_Xonly}.$a$ it
can be represented as the sum of two shifts
of the equilibrium spectrum: the reduction of the QP gap, or band gap renormalization,
due to the renormalization of the electron--electron interaction 
(blue arrow), plus the global shift due to renormalization of the
electron--hole binding energy (red arrow).
It well describes the main experimental feature,
i.e. the sharp negative peak in the TR around $3.35\ eV$.

\subsection{Complete transient reflectivity}
\label{subsec:TrAbs_theo}
We have identified two main contributions to the TR signal:
(i) the changes in the residuals of the dielectric function, which naturally extends
the concept of bleaching, photo--induced absorption and
stimulated emission, beyond the IP--RPA case;
(ii) the optical gap renormalization due to the variation of the the many--body
self--energy $\S^{cohsex}$ and the excitonic kernel $K$ of the BSE equation. This is
split in an occupations induced effect, which we find out to be almost negligible,
and a screening effect, which is instead crucial to obtain a correct
description of the main features in the experimental data.

\begin{figure}[t]
\includegraphics[width=0.45\textwidth]{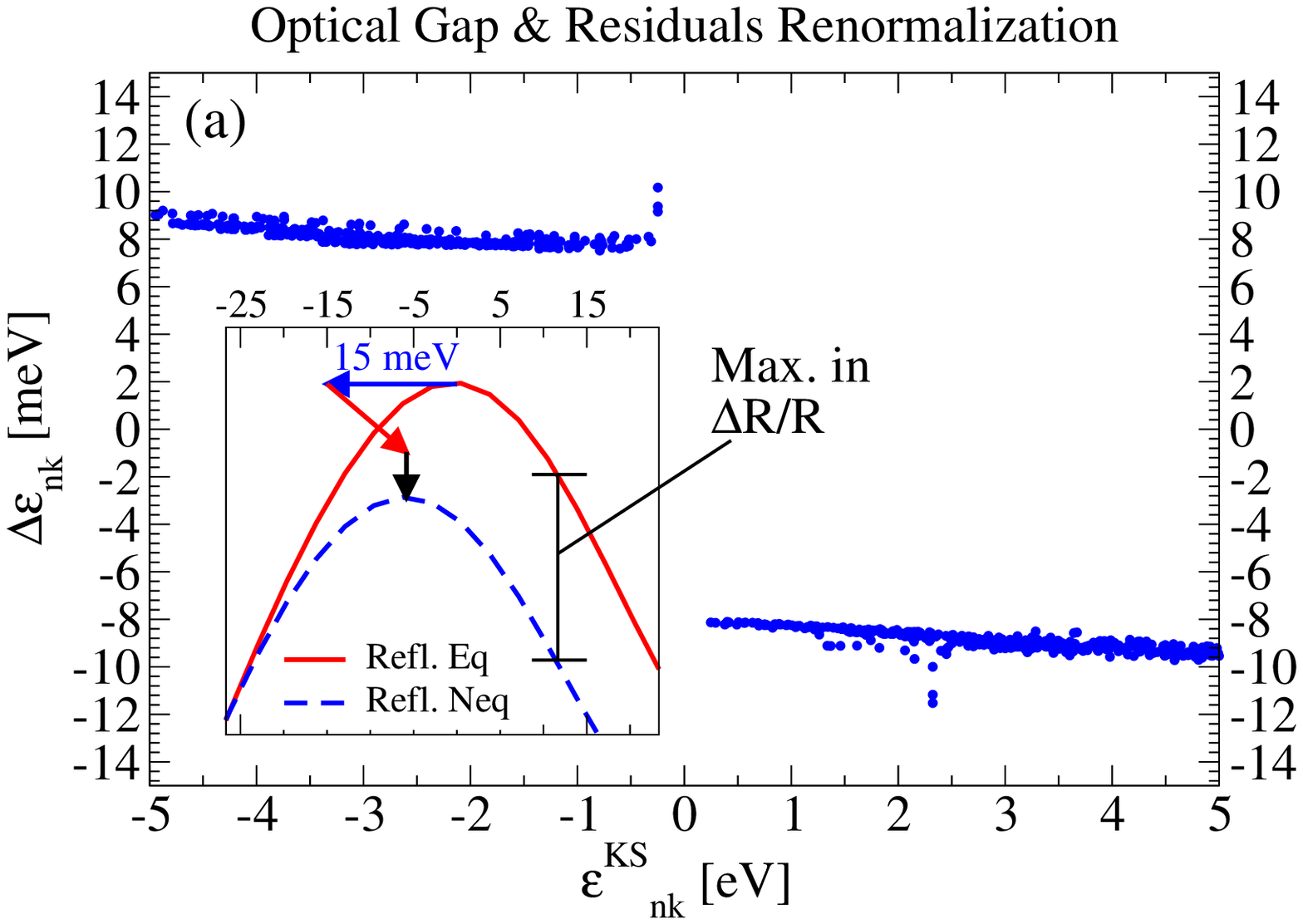}
\includegraphics[width=0.45\textwidth]{Fig8b.eps}
\caption{(color online) Renormalization of the quasi--particles energies (panel a)
         from the COHSEX self--energy including all terms.
         The change induced in the reflectivity near the main peak
         is shown in the inset (blue arrow) together with the effect due to the
         renormalization of the electron--hole interaction (red arrow) and
         the absorption bleaching of the residuals (black arrow).
         Transient reflectivity (panel b) for a
         pump intensity of $10^9\ W/cm^2$ and $\t=200\,fs$; both the variation of
         the BSE poles and the change in the residuals are considered.
         In the insets the signals at $3.45\ eV$ and at $2.1\ eV$ are
         shown as a function of the pump peak intensity. Theoretical results
         are compared with experimental data (brown circles).}
\label{fig:Transient_reflect_all}
\end{figure}

We finally compute the complete QP renormalization and 
TR signal including, together, all the effects considered in the previous sections.
In Fig.~\ref{fig:Transient_reflect_all}.$a$ the total change in the QP
corrections is the sum of the two effects represented in 
Section~\ref{subsec:TrAbs_theo_optical_gap}, with a dominant
contribution from the update of the screening.

In the same way the resulting TR (see Fig.~\ref{fig:Transient_reflect_all}.$b$)
is the sum of the effects considered in the
previous sections. The complete NEQ dielectric function is computed from
\be
\ve^{BSE}(\w,\t)= 1 -\sum_{\l\l'\in\Phi^\t}  R^*_{\l}(\t) L_{\l\l'}^{BSE}(\w,\t) R_{\l'}(\t)
\label{eq:BSE_eps_var_all}	
\text{.}
\ee
where, this time, we include the effect of NEQ occupations in the residuals
\be
R_{\l}(\t)=\sum_{l\in\Psi^\t_\qq} A^\t_{\l,l} R_{l}(\t)  \text{ where }\l\in\Phi^\t,
\ee
as well as the renormalization of the excitonic poles via $L(\w,\t)$.

The global TR signal can thus be explained
in terms of three shifts of the equilibrium reflectivity as shown in
the inset of Fig.~\ref{fig:Transient_reflect_all}.$a$.
We have a red-shift (blue arrow) and 
a second shift which is both a blue-shift and a reduction of the peak intensity (red arrow).
Last effect is a shift (black arrow) corresponding to a reduction of the
absorption intensity describing the absorption bleaching.

The computed TR spectrum well captures all the features of the experimental measure.
The only discrepancy left is a $0.15\ eV$ mismatch in the position of the negative peak.
Such error is within the common precision of first--principles calculations and can
be traced back to a combined effect of all the approximations introduced.
These approximations are essential to make these simulations possible and,
as clearly demonstrated by the deep analysis of the different contributions
performed in this work, does not weaken the predictivity power of the ab--initio approach.

We further observe that the position of the peak in $\D\mathcal{R}/\mathcal{\Rt}(\w)$
does not coincide exactly with the position of the peak in $\mathcal{\Rt}(\w)$. Such a difference
is explained in the inset of Fig.~\ref{fig:Transient_reflect_all}.$a$. The maximum distance
between the equilibrium reflectivity and the NEQ one, and thus the position of the peak in
$\D\mathcal{R}/\mathcal{\Rt}(\w)$, is shifted with respect to the peak in $\mathcal{\Rt}(\w)$
and is due to the slope of the reflectance.
This detail could be better described
only computing the smearing of the peaks and also including zero--point motion
effects in a fully ab-initio manner~\cite{Marini2008}.
Such approach however would be very demanding and beyond the 
goals of the present work.
Here we use instead a finite and constant smearing $\eta=0.1\ eV$ 
for the poles in the definition of $L(\w)$ (see Eqs.~\ref{eq:L_KS} and \ref{eq:L_BSE}).

In conclusion, beside this discrepancy, the theoretical approach captures and explains
the sharp negative peak, the negative signal in the low energy region and the positive
signal in between. The final TR signal shows a remarkably good agreement with the
experimental results, considering the complexity and totally ab--initio,
nature of the calculations.

\section{Conclusions}
\label{sec:con}
We have presented a combined theoretical and experimental study of
the transient reflectivity of bulk silicon. 

Experimentally the sample is pumped by 100-fs pulses at $\approx 3.1\ eV$,
close to the optical gap; photo--excited carriers are injected from the top of the
valence band to the conduction band thanks to the thermal broadening of the absorption
spectra.  The resulting TR spectrum is measured 200 $fs$ delay after
the pump excitation.
Theoretically, the interaction with the pump pulse is described via the time propagation of
the density matrix projected on the basis set on which the equilibrium Hamiltonian is diagonal.
The interaction with the probe pulse is then described within linear response theory computing
the non--equilibrium Bethe--Salpeter equation.

By means of this combined approach we gave a formal and accurate definition of commonly
used concepts like the photo--induced bleaching, absorption
and stimulated emission in terms of the renormalization of the optical
residuals. Thanks to the use of the Bethe--Salpeter equation these concepts are
defined within a scheme able to capture linear and non linear excitonic effects.

Another key results of the present work is the definition of {\em optical--gap renormalization}.
Such a renormalization is mainly due to the decrease of the screened
interaction, induced by the presence
of the non--equilibrium carriers created by the pump pulse. 
The optical--gap renormalization includes and extends the well known 
band--gap renormalization effect.
It is, indeed, more general as it also includes a
renormalization of the electron--hole interaction which is responsible
for an additional contribution to the transient absorption spectrum.

Combining the two effects, i.e. the residuals renormalization and the
optical--gap renormalization, 
our approach well explains the transient reflectivity signal
in silicon. The characteristic features measured experimentally
are reproduced with a simple interpretation exemplified
by three consecutive shifts of the equilibrium spectrum.

In conclusion, the present combined scheme is shown,
using silicon as a test case, to be efficient and accurate. 
The \ai\, basis makes it universal and opens the way to a new approach
able to describe ultrafast pump--probe experiments of which, this work,
is an essential step.  

Indeed, the excellent agreement with the experimental results motivates
further studies and extensions as, for example, the inclusion of relaxation and
dissipation effects and the systematic application to more systems.
The final goal is to establish a theoretical method able to describe and 
predict the outcome of pump and probe experiments in a wide class of systems.

\section*{Acknowledgements}
We acknowledge financial support by the {\em Futuro in Ricerca} grant No. RBFR12SW0J of the
Italian Ministry of Education, University and Research MIUR. DS and AM also acknowledge
the funding received from the European Union project
MaX {\em Materials design at the eXascale} H2020-EINFRA-2015-1, Grant agreement n. 676598 and
{\em Nanoscience Foundries and Fine Analysis - Europe} H2020-INFRAIA-2014-2015,
Grant agreement n. 654360. DS acknowledge Giovanni Onida for the computational time provided
on the HPC cluster ``ETSFMI'' in Milano and Paolo Salvestrini for the support on the cluster.

\bibliographystyle{apsrev4-1}
\bibliography{manuscript}

\begin{thebibliography}{72}
\bibitem [{not()}]{note_pseudoherm}%
  \BibitemOpen
  \href@noop {} {}\bibinfo {note} {The symmetrization with the sqare root of
  the occupation factors ensuares that the resonant and anti--resonant parts of
  the excitonic matrix remain hermitian, while the coupling parts remain
  symmetric. Thus the whole excitonic matrix remains the
  pseudo--hermitian.}\BibitemShut {Stop}%
\end{thebibliography}

\end{document}